\documentclass{svjour3}                     % onecolumn (standard format)
\smartqed  % flush right qed marks, e.g. at end of proof
\usepackage{graphicx}
\journalname{Experimental Astronomy}
\begin{document}

\title{Odyssey: a Solar System Mission}

\author{B.~Christophe \and 
P.H.~Andersen \and J.D.~Anderson \and 
S.~Asmar \and Ph.~B\'erio \and O.~Bertolami \and 
R.~Bingham \and F.~Bondu \and Ph.~Bouyer \and 
S.~Bremer \and J.-M.~Courty \and 
H.~Dittus \and B.~Foulon \and P.~Gil \and 
U.~Johann \and J.F.~Jordan \and B.~Kent \and 
C.~L\"ammerzahl \and A.~L\'evy \and G.~M\'etris \and 
O.~Olsen \and J.~P\`aramos \and 
J.D.~Prestage \and S.V.~Progrebenko \and 
E.~Rasel \and A.~Rathke \and S.~Reynaud \and 
B.~Rievers \and E.~Samain \and T.J.~Sumner \and 
S.~Theil \and P.~Touboul \and S.~Turyshev \and 
P.~Vrancken \and P.~Wolf \and N.~Yu
}

\authorrunning{B. Christophe and the ODYSSEY collaboration} % if too long for running head

\institute{B. Christophe \and B. Foulon \and P. Touboul \and A. L\'evy \at
ONERA, BP 72 F-92322 Chatillon Cedex \\
Tel.: +33 (0)1-46-73-49-35\\
Fax: +33 (0)1-46-73-48-24\\
\email{bruno.christophe@onera.fr}
\and P.H. Andersen \and O. Olsen 
\at University of Oslo, FFI, Norway
\and J.D. Anderson 
\at Global Aerospace, USA
\and S. Asmar \and J.F. Jordan \and J.D. Prestage
\and S. Turyshev \and N. Yu
\at NASA, JPL, Pasadena CA, USA
\and Ph. B\'erio \and F. Bondu \and G. M\'etris \and E. Samain 
\and P. Vrancken
\at Observatoire de la C\^ote d'Azur, GEMINI \& ARTEMIS, Grasse, France
\and O. Bertolami \and P. Gil \and J. P\`aramos 
\at Instituto Superior T\'ecnico, Lisboa, Portugal
\and R. Bingham \and B. Kent  
\at Rutherford Appleton Laboratory, Didcot, UK
\and Ph. Bouyer
\at Institut d'Optique Graduate School, Palaiseau, France
\and S. Bremer \and H. Dittus \and C. L\"ammerzahl  
\and B. Rievers \and S. Theil
\at ZARM, University of Bremen, Germany
\and J-M. Courty \and S. Reynaud 
\at Laboratoire Kastler Brossel, ENS, UPMC, CNRS, Paris, France
\and U. Johann \and A. Rathke 
\at Astrium, Friedrichshafen, Germany
\and S.V. Progrebenko 
\at Joint Institute for VLBI in Europe, The Netherlands
\and E. Rasel 
\at Institute for Quantum Optics, Univ. Hannover, Germany
\and T.J. Sumner
\at Imperial College, UK
\and P. Wolf
\at LNE-SYRTE, Observatoire de Paris, CNRS, UPMC, Paris, France
}

\date{Received: date / Accepted: date}

\maketitle

\begin{abstract}
The Solar System Odyssey mission uses modern-day high-precision experimental techniques to 
test the laws of fundamental physics which determine dynamics in the solar system. 
It could lead to major discoveries by using demonstrated technologies 
and could be flown within the Cosmic Vision time frame.
 
The mission proposes to perform a set of precision gravitation experiments from 
the vicinity of Earth to the outer Solar System.
Its scientific objectives can be summarized as follows: 
\textit{i)} test of the gravity force law in the Solar System up to and beyond 
the orbit of Saturn; 
\textit{ii)} precise investigation of navigation anomalies at the fly-bys; 
\textit{iii)} measurement of Eddington's parameter at occultations; 
\textit{iv)} mapping of gravity field in the outer solar system
and study of the Kuiper belt. 

To this aim, the Odyssey mission is built up on a main spacecraft, 
designed to fly up to 13 AU, with the following components: 
\textit{a)} a high-precision accelerometer, with bias-rejection system, 
measuring the deviation of the trajectory from the geodesics,
that is also giving gravitational forces;
\textit{b)} Ka-band transponders, as for Cassini, for a 
precise range and Doppler measurement up to 13 AU, with additional VLBI 
equipment;
\textit{c)} optional laser equipment, which would allow one to improve the 
range and Doppler measurement, resulting in particular in an improved
measurement (with respect to Cassini) of the Eddington's parameter. 

In this baseline concept, the main spacecraft is designed to operate beyond the 
Saturn orbit, up to 13 AU. It experiences multiple planetary fly-bys at Earth, Mars or Venus, 
and Jupiter. The cruise and fly-by phases allow the mission to achieve its baseline scientific 
objectives (\textit{i)} to \textit{iii)} in the above list). 

In addition to this baseline concept, the Odyssey mission proposes the release
of the Enigma radio-beacon at Saturn, allowing one to extend the deep space gravity test
up to at least 50 AU, while achieving the scientific objective of a mapping of gravity
field in the outer Solar System (\textit{iv)} in the above list). 

\end{abstract}

\section{Introduction}
\label{intro}
Currently, fundamental physics appears to stand at the threshold of major breakthroughs. 
Anomalies observed in the rotation curves of galaxies and in the relation between red shifts 
and luminosities of supernovae are interpreted as being due to dark matter and 
dark energy \cite{darkmatter,Riess,Perlmutter}. 
Although these dark components constitute 96\% of the energy content of the Universe, 
they have not been detected by non gravitational means to date. 
Given the immense challenge posed by this large scale behaviour of gravity, 
it is important to explore every option to explain and probe the underlying physics. 

As these anomalies could also hint at the need for gravity modifications at galactic 
or cosmological scales, it is also of extreme importance to test the laws of gravity 
at the largest possible distances. Multiple theoretical efforts, aiming to address 
the challenges above \cite{Sanders,Lue,Carroll}, suggest that a 
window to new gravitational physics could be opened at spacecraft-accessible scales 
\cite{Dittus06,RefTheory1}. 

Exploration of deep regions of the Solar System offers exciting opportunities to search 
for clues which could help to meet these challenges. 
One such clue could be the discovery made by the Pioneer 10 and 11 missions \cite{RefPioneer1}. 
The radiometric tracking of these probes at heliocentric distances between 20-70 AU 
has consistently indicated the presence of a small, anomalous Doppler drift. 
The drift was described as a constant sun ward acceleration of 
$a_P = (8.74 \pm 1.33) \times 10^{-10}$ m/s$^{2}$ for each spacecraft \cite{RefPioneer2}.
The extensive analysis of the JPL team has been published after years of cross 
checks and the presence of the anomaly has been confirmed by 
several independent analysis~\cite{Markwardt,Olsen,Levy}.
This signal has become known as the ``Pioneer Anomaly'' and its observation has stimulated 
significant efforts to find explanations in terms of systematic effects on board the 
spacecraft or in its environment (\cite{PAexplanations} and references therein). 

However, the nature of the anomaly remains unexplained today.
Further informations are expected to come out of the re-analysis process now going on
with recently recovered Pioneer data~\cite{PAdatarecovery,ISSI}.
In a context where the status of gravity theory
is challenged by the puzzles of dark matter and dark energy,
it is of importance to investigate what could be the signatures of 
the Pioneer anomaly as a long range modification of gravity
(see for example~\cite{JaekelReynaud,Moffat,Bertolami,BrunetonEsposito} 
and references therein).

The main objective of the Odyssey proposal is to study the possibility
of a dedicated mission addressing these fundamental questions.
In its baseline version, the spacecraft would fly beyond the Saturn orbit,
up to about 13 AU. 
The key novelty of the mission concept is the presence of an 
accelerometer on board, giving access to an extra observable through
the direct measurement of non geodesic forces. 
This represents a major improvement with respect to the conditions available 
at the Pioneer spacecraft, as well as at all deep space spacecraft, 
and eliminates most uncertainties which have limited navigation accuracy 
in preceding deep space missions.

During its travel to Saturn, the spacecraft experiences several planetary fly-bys 
at Earth, Mars or Venus, and Jupiter.
These fly-bys, which are needed for going to deep space while keeping a low enough
cost, are also used as ideal opportunities to gain information on the navigation 
anomalies which have been noticed at a number of recent planetary fly-bys
\cite{RefFlyby1,RefFlyby2}. 
Thanks to the availability of a precise radio-science and an accelerometer on board,
the Odyssey spacecraft would have the unique capability of monitoring 
the dynamics during planetary fly-bys and to analyse the so-called ``fly-by anomaly''.

The cruise phase could be used to test Einstein General Relativity through a 
high-accuracy measurement of Eddington's parameter $\gamma$ at solar conjunctions. 
This objective justifies the optional embarkment of laser equipment which
would lead to a measurement of $\gamma$ with an accuracy of $10^{-7}$, 
a factor of 200 beyond the current best result obtained by Cassini \cite{cassini}.
A VLBI optional equipment is also proposed for improving the resolution
of navigation on the celestial sky.

In addition to the baseline concept presented up to now, the Odyssey mission also proposes 
to consider an autonomous radio beacon probe, called Enigma \cite{Enigma}, which
would be released at Saturn, allowing one to extend the mission 
with a minimum of resources to at least 50 AU and beyond. 
This addition allows the proposal to achieve the further scientific objective of mapping 
the gravity field in the outer Solar System and, therefore, studying the Kuiper belt,
as it was initially proposed as the Zacuto concept \cite{Zacuto}. 

The efforts of the proponents has been focused on the challenge of designing a deep space mission 
within the cost of 300 Meuros. This has forced the main mission design choices 
(launcher, energy and payload options) and trade-offs in science goals. 
The payload definition emphasizes demonstrated technology, with non gravitational forces 
measured by an accelerometer upgraded from an existing qualified design and the probe 
followed by Doppler tracking and ranging built up on existing radio-science instruments. 
The present paper, drawn from the Cosmic Vision proposal \cite{ODYSSEY}, 
describes the scientific objectives, the mission profile and spacecraft design,
and gives the description of the main instruments of the mission.

\section{Scientific objectives}
\label{sec:2}
Solar System Odyssey is designed to perform a comprehensive set of gravitational tests 
in the Solar System. The mission has four major scientific objectives which are 
discussed in the following subsections~: 
\textit{i)} significantly improve the accuracy of deep space gravity test; \textit{ii)} investigate planetary fly-bys;
\textit{iii)} improve the current accuracy of the measurements of the Eddington parameter; \textit{iv)} 
map the gravity field in outer regions of the Solar System.

\subsection{Deep space gravity (DSG)}
\label{sec:2_1}

The existence of the unexpected signal known as the Pioneer anomaly is not to be doubted, 
though its nature still remains unknown. 
The deviation from expectation can be described as a constant sun ward acceleration of 
$a_P = (8.74 \pm 1.33) \times 10^{-10}$ m/s$^{2}$ for each spacecraft.
An investigation of the anomalous signal has recently been initiated \cite{ISSI}. 
The analysis of the entire record of the Pioneer spacecraft telemetry files in conjunction 
with much extended Pioneer Doppler data, has confirmed the existence of a transition region 
for $a_P$ where the effect significantly changes its magnitude:
during Pioneer 11's Saturn fly-by, $a_P$ increased by a factor of 5 compared to its pre-encounter value. 
This property is not compatible with the thermal hypothesis, one of the candidates to explain the effect. 
The transition region between 8 and 25 AU seem to hold the key to understanding the physics of the anomaly 
and, thus, it deserves a serious investigation. 

The aim of the Odyssey mission is to fly a probe beyond the Saturn orbit with modern measurement 
techniques leading to an improved accuracy and much reduced systematic errors. 
The anomaly would thus be confirmed (or disconfirmed) and new information gained on its properties 
(if confirmed), such as its direction and the constancy of its magnitude. 
Significant improvements in the mission design with respect to Pioneer probes play a role in this 
information gain.

A key novelty of the proposed mission is the presence of an accelerometer on board, 
giving information on the deviation of the probe from geodesic motion. 
This information was not available on the Pioneer spacecraft and their navigation was relying on a 
complex processing of the Doppler data, the only type of data available for the two vehicles. 
The present proposal takes advantage of accelerometer data, which directly measure non-gravitational 
forces such as the thermal recoil force generated on-board or the environment-induced forces with 
origin outside the spacecraft, namely the solar pressure and interplanetary drag force. 
It would also allow one to perform measurements at earlier phases of the mission, which is 
critical for measuring the onset of the anomaly.

Another important advantage of the proposed experiment is the use of modern radio-techniques that result 
on a much improved measurement accuracy and better control of the systematic errors. 
The availability of a range observable significantly reduces the uncertainty in the determination of the 
heliocentric distances of the probe \cite{JaekelReynaud}. 
Meanwhile, the VLBI techniques give new angular observables related to transverse motion with respect 
to the line of sight.

This unique combination of measurement techniques will allow Odyssey to investigate gravity 
in the Solar System as its primary objective up to 13 AU. The radio-beacon Enigma will extend the deep space gravity test up to at least 50 AU, without accelerometer on board, but with accurate non gravitational acceleration models. Specifically, this mission is going to 
accomplish the following: 
\begin{itemize}
\item Investigate the gravitational inverse-square law in the Solar System at distances 
from 1 AU to at least 50 AU with a goal of 80 AU. 
To accomplish this objective, the mission will search for dynamic anomalies in the motion 
of the spacecraft with an accuracy at the level of $4 \times 10^{-11}$ m/s$^2$ 
(respectively $8 \times 10^{-12}$ m/s$^2$), for the main spacecraft up to 13 AU 
(respectively for the Enigma radio beacon probe from 10 AU to 50 AU); 
if any anomaly is found, the mission will establish the true direction of the anomaly. 
\item Study the onset of the anomaly, in particular at planetary encounters 
(e.g. the effect seen during Pioneer 11's Saturn fly-by). 
\item Assess the temporal evolution of the anomaly with a goal of determining the long-term 
temporal behaviour of the anomaly. 
\end{itemize}

The most critical requirement needed to satisfy the objective above is to perform the Doppler 
acceleration measurement with the quoted accuracy. 
The basic observable quantities for these experiments are:
\begin{itemize}
\item The range and range-rate between the ground stations and the spacecraft, 
eventually after removing the effects of the plasma along the path by means of a 
multi-frequency link in X and Ka-bands.
\item The non-gravitational perturbations acting on the spacecraft, 
measured by the on-board accelerometer.
\item The absolute attitude of the spacecraft, in a stellar frame of reference, 
by means of a star tracker.
\item For the radio-beacon, the precise determination of solar radiation pressure 
and minimization of on board distortions.
\end{itemize}

In addition to the standard line-of-sight radiometric tracking techniques, 
the Odyssey mission will use VLBI data. 
This additional data will allow one to significantly constrain the uncertainty of the 
spacecraft position on the sky by determining its angular position with respect 
to the set of pre-selected quasars. 
As information at the same level of accuracy will be provided by the accelerometer, 
the VLBI information will be used for the following goals:
\begin{itemize}
\item checking, from time to time, the quality of the trajectory reconstruction;
\item acquiring new information on the direction of the anomaly: towards the Sun? 
towards the Earth? opposite the direction of motion? 
\end{itemize}

The different navigation measurement (acceleration, ranging, Doppler, VLBI) 
shall be mixed to provide the best navigation of the spacecraft and to deduce 
the gravity field seen during its trajectory.

\subsection{Investigation of planetary fly-bys}
\label{sec:2_2}

In several missions using Earth gravity assistance, it has been noticed that, after the fly-by, 
the spacecraft possessed a velocity larger than that calculated from the precisely measured 
initial conditions and the now well known properties of the Earth gravity field 
\cite{RefFlyby1,RefFlyby2}. 
The anomalous additional velocities are on the order of 1 cm/s and are significantly larger 
than the measurement accuracy: the gravity field of the Earth, atmospheric drag, 
charging and Earth tides have been shown to result in uncertainties well below the measured effect. 

Odyssey uses the opportunity of multiple planetary fly-bys during the travel to Saturn. 
It takes benefit from this situation to investigate fly-bys with well defined initial conditions, 
measurement of non gravitational forces by the accelerometer, and accurate tracking techniques. 
The accelerometric measurement is available during the whole duration of the fly-by, 
including the typical DSN black-out period. 
Not only fly-bys at Earth, but also fly-bys at other planets, will be studied in order to 
decide whether the effect is universal or just related to the Earth. 
Completion of this objective will provide a better understanding of satellite dynamics 
and thus to have a significant impact on improved navigation techniques. 
This is true whether or not this effect can be interpreted within conventional physics. 

For an accurate characterization of the fly-by, the velocities will be measured by the 
radio-tracking system before and after the fly-by with a precision of 10 $\mu$m/s. 
This is less than 1\% of the typical anomalous increment $\Delta V$ registered on 
presently available data. 
VLBI data will also be used for a better determination of the spacecraft position 
at the asymptotes of its hyperbolic path. 
The performance of the accelerometer, as shall be discussed, with 3-dimensional acceleration 
resolution at the level of $4 \times 10^{-11}$ m/s$^2$, is largely sufficient to achieve 
the same accuracy on the integration of the non gravitational force during the fly-by. 
As the gravitational acceleration is known with an accuracy better than needed, 
the instruments on board Odyssey will unambiguously confirm (or deny) the existence of a 
discrepancy between the expected and measured $\Delta V$, the major improvement being 
the resolution of the uncertainty on non gravitational forces (whatever their origin). 

A dynamical range of the accelerometer of 20 $\mu$m/s$^2$ is sufficient to reach this goal 
of a closest approach to Earth as low as 600 km.
This figure has been calculated for a cross section of 20 m$^2$, corresponding to fully 
deployed solar panels, a spacecraft mass of 420 kg and the less favorable assumption for 
the atmospheric drag model.

\subsection{Solar conjunction}
\label{sec:2_3}

The Eddington's parameter $\gamma$, whose value in General Relativity is unity, is a 
fundamental parameter in most tests of relativistic gravity. 
In fact, $(1- \gamma )$ measures the deviation from General Relativity of competing theories. 
This deviation has been shown to be smaller than $2 \times 10^{-5}$ by the Cassini relativity 
experiment performed at solar conjunctions \cite{cassini}. 
But there exist theoretical models~\cite{Damour1,Damour2,Damour3} which suggest that this deviation 
might have a natural value in the range $10^{-6}$ as a consequence of a damping of the scalar 
contribution to gravity during cosmological evolution.

This is clearly a motivation for repeating the Cassini relativity experiment.
If using the same radio-science as Cassini, the experiment will confirm the results of Cassini 
with the advantage of an accelerometer on board to measure the non-gravitational acceleration. 
As this information was not available with Cassini, it was reconstructed as one parameter of the best fit. 
In Odyssey, this value will be measured by the accelerometer, leading to an improved version of the 
Cassini relativity experiment.

A largely improved accuracy could be attained if laser ranging equipment could be embarked on board. 
Odyssey could thus determine the parameter $(1- \gamma )$ at a level of accuracy below $10^{-7}$, 
which would provide new crucial information on scalar-tensor theories of gravity at their fascinating 
interface with theories of cosmological evolution. 
This is the main motivation for the up-scaling laser option, which would also improve range
and Doppler performances. 
The improved performances brought by laser tracking are discussed in the SAGAS proposal~\cite{SAGAS}. 

\subsection{Study of the outer Solar System}
\label{sec:2_4}

In the context of the modern fundamental physics , it is important to use any relevant information 
which can be drawn from a mission like Odyssey for a complete and unambiguous mapping of the 
gravity field in the outer Solar System. 
An under-estimated mass of the yet unexplored Kuiper belt could produce an anomalous acceleration. 
The variation of a gravitational acceleration due to Kuiper objects will be higher than 1 pm/s$^2$ 
over a distance of 20 to 80 AU from the sun \cite{Zacuto}. 

This range of distance will be covered by the Enigma radio-beacon. 
If the orbit reconstruction on the Enigma radio-beacon can be mastered at this level of accuracy, 
then the mission could result in a precise mapping of the gravity field in the outer Solar System. 
In particular, it would, for the first time, be able to discriminate between the different models 
which have been proposed for the spatial distribution of the Kuiper belt.

\subsection{Scientific scenario}

The preceding description of scientific objectives of Odyssey shows a set of options.  

The baseline version is a probe flown to deep space (up to 13 AU), with an on board accelerometer 
and radio-metric tracking. 
This unique instrumental combination allows for a gravity test with an unprecedented accuracy on 
the journey from Earth to deep space, including several planetary fly-bys. 
The main improvement option releases Enigma radio beacon able to fly to much larger heliocentric 
distances (at least up to 50 AU, and maybe up to 80 AU). 

Two other important improvements are the VLBI and laser ranging options. 
The VLBI option improves accuracy of the angular motion of the probe on the celestial sky while 
the laser ranging option improves the accuracy of tracking, 
with a particularly impressive effect on solar conjunction experiments. 
The technological readiness of laser ranging is smaller than that of the radio-science, 
but its presence in the mission, if feasible, would bring a large improvement for the measurement of $\gamma$. 

\section{Mission profile and spacecraft}
\label{sec:3}

\subsection{Mission profile}
\label{sec:3_1}

For the currently foreseen spacecraft of 420 kg mass at Earth escape, a launch with Vega is feasible. 
So, it is mandatory to first inject the spacecraft into Low Earth Orbit (LEO) 
and from there achieve Earth escape by means of a propulsion module (PM). 
The bi-propellant propulsion module is derived from LisaPathfinder, already in the advanced design phase. 
After injection into LEO, escape from Earth's attraction is achieved by a apogee raising sequence. 
The duration of the apogee raising sequence will be on the order of 1 month.

In order to further increase escape performance, a lunar gravity assist (LGA) could be considered. 
This manoeuvre will be demonstrated by the BepiColombo spacecraft, and hence will be an established 
practice by the launch date of Odyssey. 
Assuming a deflection angle of 40 degrees, a $\Delta V$ gain of 400 m/s is achievable. 
A powered lunar fly-by using the propulsion module could be also considered, in order to further improve performance.
By this escape strategy, hyperbolic excess velocities on the order of 5 km/s are achievable for 
the current Odyssey spacecraft. This is more than sufficient for the planned interplanetary trajectory.

An important requirement of the interplanetary trajectory is to have several planetary gravity assists. 
On one hand, these serve the science goal of precision tracking of the Odyssey spacecraft during fly-bys. 
On the other hand they enable a highly energetic, hyperbolic trajectory toward the outer Solar System,
while staying within a limited cost. Currently, two options are considered to accomplish these goals.

The well-known Earth-Venus-Earth-Earth-Jupiter-Saturn (EVEEJS) series of fly-bys and the 
Earth-Mars-Earth-Earth-Jupiter-Saturn (EMEEJS) sequence both meet our goals. 
The latter has slightly higher $\Delta V$ requirements, but allows for a less demanding thermal 
control by avoiding the low heliocentric distance of Venus.
Both trajectories lead to a hyperbolic escape velocity of 23.5 km/s, which is mainly determined by 
the Saturn fly-by.
The release of Enigma radio-beacon occurs during the Saturn fly-by, where the non-gravitational 
accelerations on the radio-beacon are minimised with respect to an earlier release.

\subsection{Spacecraft}
\label{sec:3_2}

The spacecraft is designed for the complete scenario, in order to maximise the scientific return: 
all the instruments described in the next chapter are included, even if some instruments are slightly 
redundant (laser tracking and Ka transponder). 
This capability to design such a spacecraft, compatible with a launch with VEGA, highlights that some 
margins exist which should be finely analysed during the assessment study.

The spacecraft design responds to the requirements of the payload and the challenges of the trajectory. 
The accelerometer is accommodated in the middle of the spacecraft, close to the centre of mass. 
In order to minimise the shift in the centre of mass during the mission, the propellant is distributed 
between two tanks which are kept at a roughly equal levels of fill throughout the journey. 
The configuration is characterised by:
\begin{itemize}
\item A cube-shaped spacecraft  bus of 1 m edge length
\item Four deployable rigid solar arrays of 5 m$^2$ each. 
\item Radiators behind louvers in the shadow of solar arrays
\item A fixed High Gain Antenna (HGA) of 1 m diameter (option 0.5 m)
\end{itemize}

The total wet mass of the spacecraft is 422 kg, with 50 kg of propellant for cruise and attitude manoeuvres. 
The propulsion module mass is  based on the LISAPathfinder PM mass budget and the EUROSTAR tank mass 
specifications, with a dry mass of 226 kg. For Earth Escape, 1600 kg of propellant is needed, 
leading to a total mass at launch of 2240 kg. 

Despite its journey slightly beyond the orbit of Saturn, the Odyssey spacecraft will rely solely 
on solar arrays for power generation. During the journey the available power will change over 
two orders of magnitude between the closest point to the Sun 
- the Venus fly-by (in case the EVEEJS scenario is chosen) - 
and the end of the spacecraft mission. 
In order to cope with these changing conditions, the power consumption will be gradually reduced 
towards the end of the mission by selectively switching off certain equipment.
At 7.5 AU the laser receiver and the clock will be put out of service 
(at this stage the relativity experiments for which the laser was needed will have been conducted). 
Shortly after, the reaction wheels will be switched off and the spacecraft will be put into a spin stabilised mode.

With these power saving techniques, the mission will continue until around 13 AU, where power will become 
insufficient for the operation of the thruster heaters. 
After the end of the nominal mission, it may be possible to continue for a while
with a reduced number of thrusters and limited attitude control. 
In the meantime, the autonomous radio beacon probe Enigma will have been released (at Saturn fly-by), 
allowing the mission to be extended to a distance far beyond the baseline spacecraft capabilities. 
It will thus appear natural to scale down its operations and stop them 
not too long after the Saturn fly-by without a significant loss of scientific return.

\section{Instruments}
\label{sec:4}

The present section is devoted to a discussion of the main instruments
of the Odyssey mission. 

\subsection{$\mu$STAR Accelerometer}

The accelerometer package is composed of an electrostatic accelerometer, with its electronics 
and a bias compensation system. 
The core of the instrument is an electrostatic accelerometer based on Onera expertise in 
the field of accelerometry and gravimetry (CHAMP \cite{RefCHAMP}, GRACE, GOCE missions \cite{RefAcc}). 
Ready-to-fly technology is used with original improvements aimed at reducing power consumption, size and weight. 

The bias compensation system consists in a flip mechanism which allows a 180$^\circ$  rotation of the accelerometer 
to be carried out at regularly spaced times. 
The flip allows the calibration of the instrument bias along 2 directions, by comparing the acceleration 
measurement in the two positions. 
The instrument can thus measure all the non-gravitational forces with an accuracy of 
$4 \times 10^{-11}$ m/s$^2$, which is largely sufficient to confirm/disprove the presence of an anomaly 
at the Pioneer level of $8 \times 10^{-10}$ m/s$^2$. 

The mechanical core of the accelerometer is composed of a cubic proof-mass, with 3 pairs of similar 
electrode plates, each pair controlling two degrees of freedom. 
The base plate and the housing ensure a good vacuum. 
The control of the proof-mass is performed by low consumption analogue functions. 
The output of the accelerometer, which is the voltage applied on the electrode to control the proof-mass, 
is sent to an Interface Control Unit, which can be shared with other instruments of the spacecraft. 

The expected noise level is $10^{-11}$ m/s$^2$ rms, integrated over one day, 
along one sensitive axis with a light proof-mass (18 g), 
when considering a passive thermal stability of 0.1°C at 0.1 mHz, 
assuming a data rate of 300 bits/10 s.
In this configuration, the accelerometer range is $2 \times 10^{-5}$ m/s$^2$ and the maximum bias 
level before calibration is $2 \times 10^{-6}$ m/s$^2$, with a thermal stability of 
$7 \times 10^{-10}$ m/s$^2$/K with respect to the temperature of the mechanical sensor 
(due mainly to the stiffness variation of the gold wire). 
A stability of 10 mK is sufficient to have a variation less than $10^{-11}$ m/s$^2$.
The budget of the instrument package is a volume of 3 l, a mass of 3 kg and a maximal 
consumption of 3 W. 

The calibration system, with stepping motor, is added in order to rotate the complete 
accelerometer by 180 degrees, around the vertical axis of the instrument. 
The bias along both horizontal axes of the accelerometer, the ultra-sensitive axes, 
will be calibrated by comparing the measurement before and after tilting 
(these ultra-sensitive axes are in the orbit plane, one of them being aligned with the LOS axis).
The alignment accuracy of the calibration system between the two positions should be less than 
0.25 mrad in all directions: this allows an error less than $2.90 \times 10^{-11}$ m/s$^2$ 
in the worst case at 2 AU, and less than $4.63 \times 10^{-12}$ m/s$^2$ in the worst case at 5 AU.
The momentum compensation will be done by counter-rotating masses.

The main challenge is the integration of the accelerometer inside the spacecraft. 
Ideally, the accelerometer should be placed at the centre of gravity of the spacecraft, 
in order not to perturb the measurement by the spacecraft angular rate and angular acceleration. 
A preliminary analysis shows that the objective of an accuracy of $10^{-11}$  m/s$^2$ 
is feasible with a centring or knowledge of the accelerometer position at a level of 0.5 mm, 
and an angular acceleration known at better than $10^{-8}$ rad/s$^2$.

An accurate knowledge of the accelerometer position with respect to the centre of 
gravity of the spacecraft, and an accurate measurement of the spacecraft angular rate and 
acceleration during the mission would allow for a relaxation of this requirement. 
The calibration of the decentring could be done in-flight, with a specific angular acceleration 
signal and the post-processing of the accelerometer outputs. 
It should also be possible to modify the centre of gravity of the spacecraft
with a mass trim mechanism. 

The noise performance and the bias stability depend also on the thermal stability 
at the accelerometer location: a thermal stability of 0.1°C at 0.1 mHz is considered. 
For the bias drift, the requirement of an accuracy of $10^{-11}$ m/s$^2$ leads to a 
requirement of a long-term thermal stability of 10 mK. 
Again, the possibility of a correction should be considered, if the thermal bias drift 
can be calibrated and the temperature measured at this level. 

\subsection{Multi-frequency radio-links}

The key to achieving the target accuracies is the use of a multi-frequency radio link, 
which takes advantage of the dispersion of the plasma to calibrate the noise. 
In this scheme, two coherent signals are transmitted and received simultaneously at 
X- and Ka-band, and a third mixed link is added in which the X-band uplink is transponded at Ka-band. 
The third link is required by the different (and unfortunately fixed) transponding ratios 
at X- and Ka-band. One of the two channels (either X or Ka) would be the standard radio link for 
commands and telemetry; the others are specific to radio science and the associated instrumentation 
is therefore considered as payload. 
This approach has been used for the Cassini \cite{cassini} and BepiColombo \cite{RefBepi} spacecraft, 
but the calibration of the plasma noise is only possible for Doppler observables and is baselined for Odyssey.

The crucial component of the plasma noise cancellation system is a Ka-band radio link 
(34 GHz uplink, 32.5 GHz downlink) originally developed for Cassini radio science experiments. 
The 34-m DSN station in California (designated DSS-25) underwent significant upgrades to meet 
stringent radio science requirements for precision Doppler investigations, including the 
separation of the transmitting and receiving feeds to account for aberration effects.
On board the Odyssey spacecraft, the key instrument is a Ka-band Transponder (KaT) that 
coherently re-transmits the uplink signal back to the ground station. 

The instrument will receive a Ka-band uplink signal from the Deep Space Network and generate a 
downlink signal coherent with the uplink, maintaining the excellent phase stability of the uplink. 
The downlink, when received on Earth, is considered high quality precision Doppler. 
The ultimate use of the KaT involves simultaneous reception of X-band data from the spacecraft's 
telecommunication transponder and two processes of calibration, one for complete removal of the 
dispersive noise and a second to remove the effect of the Earth's ionosphere.

The key for a good accuracy of tracking data is a very high phase stability of the 
microwave signal by means of a two-way radio link. 
In a typical deep space configuration, the ground station generates the uplink signal 
using a H-maser, capable of delivering a frequency stable to one part in $10^{15}$ over time scales of 1000 s. 
This signal is converted up to X and Ka-band (7.2 and 34 GHz), amplified and transmitted to the spacecraft. 
For ranging measurements, the carrier is modulated with a suitable tone with a bandwidth as large as 50 MHz 

The requirements on the tracking system for frequency stability (expressed in term of Allan deviation) 
and ranging accuracy may be summarised as 
$\sigma_y= 10^{-14}$ for integration times between $10^3$ and $10^4$ s,
and $\sigma_\rho=1$ m two-way.
In terms of a more familiar velocity measurement, the first figure corresponds to a two-way range-rate error
of $3 \times 10^{-6}$ m/s. 

\subsection{VLBI otpion}

The main goal of the Very Large Baseline Interferometer (VLBI) up-scaling option
is to determine the direction of the residual kinematical acceleration of the spacecraft
using accurate measurements of its angular position, provided the radial acceleration is
measured by the ranging and Doppler tracking instruments. 
The experiment would work for both the spacecraft during its cruise phase up to 13 AU distance, 
and the Enigma radio beacon probe to 50 AU distance and beyond. 
The geometric sensitivity of VLBI observations (70 $\mu$as for a single 10-hour observational run) 
determines the direction of the residual acceleration with an accuracy sufficient to distinguish 
between Sun-centric or alternative models of the residual acting force.

The ground segment of the VLBI experiment consists of a number of radio telescopes observing the 
spacecraft and natural radio sources in group delay or phase referencing mode.  
As a first scenario, we propose to use the group delay mode, standard for astrometric and geodetic 
VLBI observations used to create the current celestial reference frame \cite{RefVLIBI}.

At a given power and antenna gain of the spacecraft transmitter, the signal to noise ratio 
for 2.5 W carrier line and 0.25 W additional tones, coherent with the carrier, 
transmitted from 10 AU distance and with an integration time of 120 s, 
will be more than 100 when received by 25 m antennas with SEFD = 300-400 Jy. 
A tone separation of 300 MHz will provide a better than 40 ps group delay stochastic error 
per single baseline, per integration time. With the array of ~10 telescopes 
and a mean baseline of 3000 km, ~50 $\mu$as stochastic angular accuracy will be achieved 
after solving for all baselines and averaging over ~20 scans. 
Nodding the array between the spacecraft and ~100 reference sources over ~10 hours will 
lock the spacecraft positional measurements to the Celestial Reference Frame with ~50 $\mu$as accuracy. 
Reference sources will be observed in dual-frequency S/X mode to resolve the ionospheric contribution. 
Observing a near by (1-2 degrees separation from the spacecraft) source immediately before or after 
each spacecraft scan will allow transfer of the reference ionospheric delay to the spacecraft measurement, 
so enabling ~ 70 $\mu$as (or 500 m at 10 AU) accurate positioning of the spacecraft in the ICRF/SSBC frame. 

Seven to eight observations over a 9 month time base will determine the residual lateral acceleration 
with 10 pm/s$^2$ accuracy, or vectorize the 1 nm/s$^2$ radial acceleration with 0.01 radian accuracy, 
discriminating in this way between Sun-centric or Earth-centric models. 

Unlike the spacecraft VLBI experiment, the Enigma radio-beacon probe VLBI experiment will face the major 
problem of signal weakness. A similar observing strategy can be applied, but will require the inclusion 
in the array of at least one ~100 m and SEFD 15-20 Jy telescope and several 30-40 m class telescopes 
with SEFD 150-250 Jy, extending the array to 12-16 telescopes, and an increase of the coherent integration 
time on tone detection to 1000 s, to achieve 100 $\mu$as positioning accuracy for a single observational run. 
With 10-15 observations over a time base of 2-3 years, the same vectorization accuracy, as in the case of 
the spacecraft at 10 AU distance, will be achieved relative to the current Sun-radio beacon-Earth angle. 
Because no ranging for the Enigma probe is foreseen, it will rely on the Doppler tracking to determine the 
radial acceleration.
 
Space segment parameters used for VLBI accuracy estimates are listed in Table ~\ref{tab:VLBI}.
\begin{table}[h]
	\caption{ Space segment parameters required for VLBI}
	\centering
	\label{tab:VLBI}       % Give a unique label
	\begin{tabular}{|c|c|c|}
		\hline
		 Parameter&  Main craft & Enigma \\
		\hline
		 TX antenna gain  & +32 dBi & +3 dBi \\
		\hline
		 TX power in carrier & 2.5 & 25 W \\
		\hline
		TX power in tones &	$4 \times 0.25$ W & $4 \times 2.5$ W \\
		\hline
		 Tone separation & 300 MHz & 300 MHz \\
		\hline
		 USO stability & $< 10^{-13}$ @ 100 s & $< 10^{-13}$ @ 100 s \\
		\hline
	\end{tabular}
\end{table}

%\begin{table}[h]
%	\caption{ Space segment parameters required for VLBI}
%	\centering
%	\label{tab:VLBI}       % Give a unique label
%	\begin{tabular}{|c|c|c|}
%		\hline\noalign{\smallskip}
%		 Parameter&  Main craft & Enigma \\
%		\hline\noalign{\smallskip}
%		 TX antenna gain  & +32 dBi & +3 dBi \\
%		\hline\noalign{\smallskip}
%		 TX power in carrier & 2.5 & 25 W \\
%		\hline\noalign{\smallskip}
%		TX power in tones &	$4 \times 0.25$ W & $4 \times 2.5$ W \\
%		\hline\noalign{\smallskip}
%		 Tone separation & 300 MHz & 300 MHz \\
%		\hline\noalign{\smallskip}
%		 USO stability & $< 10^{-13}$ @ 100 s & $< 10^{-13}$ @ 100 s \\
%		\noalign{\smallskip}\hline
%	\end{tabular}
%\end{table}

\subsection{Laser tracking option}

The laser option, called TIPO (T\'el\'em\'etrie Inter Plan\'etaire Optique \cite{RefTIPO1}), 
is a one-way laser tracking project derived from satellite and lunar laser ranging (SLR/LLR) 
and optical time transfer T2L2 \cite{RefTIPO2}. 
The TIPO principle is based on the emission of laser pulses from an Earth based station towards 
the spacecraft. These pulses are timed in the respective timescales at departure on Earth and 
upon arrival on the spacecraft. 
The propagation time and the respective distance between Earth and spacecraft are derived from 
the difference of the dates of departure and arrival. 
Observations, implying a network of Earth laser stations on a global scale, 
allow for measuring the angular position of the spacecraft and thus the position in the plane 
perpendicular to the line of sight. This one-way laser ranging permits distance measurements 
on a Solar System scale (up to billions of km) as the one-way link budget varies only with the 
square of the distance, contrary to the power of four for usual two-way laser ranging.

The TIPO project is composed of a ground and a space segment.
The ground segment is a high-fidelity laser ranging station, working solely as an emitter 
(as the Satellite and Lunar Laser Ranging station 'MeO' of OCA in Grasse, France).
The space segment is composed of an optical and an electronics subsystem.
The optical subsystem comprises a rather compact telescope in order to collect the incoming 
laser pulses, a spectral filter for noise reduction and a linear, high bandwidth photon 
detection system (APD or MCP photo multiplier) yielding good timing characteristics. 
Considering maximum distance of 7.5 AU, a telescope with a diameter of 25 cm should be sufficient.
The electronics include a timer assembly, control electronics and a frequency synthesis. 
In addition to its normal function (coarse and fine timing) the timer assembly features 
a real-time temporal filter for signal to noise ratio (SNR) enhancement.
For this purpose, the laser ground station emits the laser pulses with an imposed phase 
relationship (a temporal code). After a first acquisition ($e.g.$ over 1000s) by the coarse timer 
(resolution 10ns), the electronics convolute the generated events applying the code used by the 
ground station, and thus yield a prediction for the upcoming events. 
By this means the raw SNR (which is on the order of 0.1 to 1 for reasonable 
telescope sizes) may be raised by one to two orders of magnitude to a satisfying signal confidence level.

The main mass and volume driver in the TIPO instruments is the telescope diameter, 
which in turn is imposed by the mission distance. 
Table ~\ref{tab:TIPO} summarises the estimated resources for the subsystems.

\begin{table}[h]
	\caption{Budget for TIPO instrument}
	\centering
	\label{tab:TIPO}       % Give a unique label
	\begin{tabular}{|c|c|c|c|}
		\hline
		 & Mass & Volume & Power \\
		 & [kg] & [l] & [W]  \\
		\hline
		 Optical subsystem  
			& 8 & 16 & - \\
		 (telescope, Ø 25cm) 
			&   &   &   \\
		\hline
		 Electronics subsystem  
			& 4 & 12 & 12 \\
		\hline
		 Clock 
			& 2 & 2 & 14 \\
		\hline
		 Total 
			& 14 & 30 & 26 \\
		\hline
	\end{tabular}
\end{table}

%\begin{table}[h]
%	\caption{Budget for TIPO instrument}
%	\centering
%	\label{tab:TIPO}       % Give a unique label
%	\begin{tabular}{|c|c|c|c|}
%		\hline\noalign{\smallskip}
%		 & Mass & Volume & Power \\
%		 & [kg] & [l] & [W]  \\
%		\hline\noalign{\smallskip}
%		 Optical subsystem  
%			& 8 & 16 & - \\
%		 (telescope, Ø 25cm) 
%			&   &   &   \\
%		\hline\noalign{\smallskip}
%		 Electronics subsystem  
%			& 4 & 12 & 12 \\
%		\hline\noalign{\smallskip}
%		 Clock 
%			& 2 & 2 & 14 \\
%		\hline\noalign{\smallskip}
%		 Total 
%			& 14 & 30 & 26 \\
%		\noalign{\smallskip}\hline
%	\end{tabular}
%\end{table}

Considering an on-board clock that is initially not synchronised with an Earth temporal reference frame 
(i.e. switch-on in space) and whose frequency is unknown (poor exactitude), 
TIPO nevertheless allows differential distance measurements and the determination of the anomaly, 
afflicted with a scale factor that is a function of the clock frequency offset.
A timing precision of about 10 ps would allow for an initial distance measurement on the millimetre level.
A clock like the space-proven Perkin-Elmer Rubidium, with a frequency flicker noise floor of 
$10^{-14}$ around $10^4$ s, would degrade the precision to a value of about 100 ps, 
considering a measurement rate of 10 Hz (classical laser station).

\subsection{Enigma radio-beacon option}

The radio beacon probe will begin an autonomous mission after Odyssey has reached a distance 
from the Sun such that the solar cell output decreases below an operational level \cite{RefEnigma}. 
The radio beacon probe will then continue to the outer Solar System on its own trajectory. 
Odyssey spacecraft carries Enigma to its launch point at 10 AU (after Saturn fly-by) 
where it is injected into a solar escape trajectory.
The total mass of Enigma is 54 kg, including margin.

The key concepts which enable Enigma to carry on the mission without resorting to 
big RTG (Radio-isotope Thermal Generator) power supplies, in contrast to all past 
classical missions to the outer Solar System (with the exception of Rosetta), is:
\begin{itemize}
\item Hibernation mode most of the time to charge batteries
\item Short periodical burst modes to perform radio science
\item Built in very low power ($<$2 W$_{el}$ total) RTG for power and thermal control (RHU-type)
\item Minimum of diagnostic sensors, but a very good model of radiation pressure interaction due to dedicated design
\item Minimised and balanced on board distortions
\item Passive spin stabilisation
\item Low gain radio antenna supporting downlink up to 50 AU (extended, beyond 80 AU) for Doppler range rate measurement
\end{itemize}

In particular, Enigma is designed to allow precision tracking of the geodetic motion 
of a probe in the far Solar System range scale with a minimum of instrumentation, reducing 
resource needs and thus avoiding challenging design drivers. 
It focuses on essential measurements to unambiguously verify and characterise the gravity 
by radio range rate (Doppler) and lateral tracking (ranging optional) with the alternative 
DSN stations, Greenbank and Effelsberg. A one-way link is sufficient in case a precise on-board 
clock is available in time (few $10^{-14}$ in $10^6$ s). Alternatively, the link budget also allows 
two-way coherent radio science if the large dish stations and high transmitter power are used (1 kW). 
Baseline is X-band, but K-band is a viable alternative. 
Requirements on plasma effects compensation are moderate.

The probe acceleration shall be measured to an accuracy of $8.74 \times 10^{-12}$ m/s$^2$ which 
corresponds to a value of 1\% of the Pioneer Anomaly. In order to accomplish this goal the local 
solar constant has to be measured and the solar pressure acceleration on the probe has to be modelled, 
both with adequate precision. The required precision is realised by a dedicated design of the surface 
properties, and the calibration of its radiation properties on ground as well as in space, taking into 
account ageing effects. Furthermore, perturbations generated on board (thermal dissipation power and 
uniformity, out-gassing) are minimised by design, and further by spin averaging their effects along the 
radio science line of sight below the required measurement accuracy.

\section{Conclusions}
\label{sec:5}

The Odyssey mission relies on several European institutes, specialized in fundamental physics, 
in precise navigation instrumentation, or in accurate orbit determination.
American partners, largely engaged in the first deep space gravity experiment (Pioneer 10 and 11), 
collaborate on this mission.
These institutes have already collaborated in the frame of ISSI, for the Pioneer Data analysis. 
This collaboration has led to development of new tools, in particular for the accurate orbit determination, 
which will support the science operations of the Odyssey mission \cite{ISSI}.

The proposed instruments have high technological readiness level, allowing a limited cost mission:
\begin{itemize}
\item The accelerometer $\mu$STAR is based on the technology heritage of Onera for ultra-precise 
accelerometry (CHAMP, GRACE, GOCE...). 
The development of the innovative calibration system should allow ESA to have a bias-free 
accelerometer with off-the-shelf technologies.
\item Europe is already at the forefront of Ka-band space telecommunications: ASI and Alenia Spazio 
have already provided a Ka-Ka frequency transponder for the radio science experiments of the Cassini 
mission and this has led to the current best test of general relativity.
\item The ground-based VLBI network at its current level is ready for spacecraft tracking with the 
specified accuracy. Mission specific software developments and test observations are required. 
Major upgrades of the global VLBI network, which will improve its accuracy by a factor of 2-3, 
are planned independently of the proposal.
\item The laser tracking option TIPO has already been proposed for the joint NASA/CNES Mars mission MSRO. 
It relies on an important technological heritage, derived from satellite and lunar laser ranging 
(SLR/LLR) and the instrument T2L2 which is presently integrated in the Jason 2 satellite (launch in 2008). 
\item The ENIGMA concept is an ambitious up-scaling option, feasible within the Cosmic Vision 
time frame. The clock stability for a one–way downlink needs dedicated improvement 
as well as the optional micro-accelerometers at the needed accuracy. 
ENIGMA opens the way to gravity tests at extremely large distances.
\end{itemize}

Odyssey offers an outstanding opportunity for communication and outreach with a very large public impact. 
The relativistic conception of space-time, and the experiments and observations which test this theory, 
raise fascinating questions. 
The fact that observations fit General Relativity only when assuming that 96\% of the Universe is 
composed of dark components (dark matter and dark energy), the nature of which has not yet been 
given a satisfactory explanation in fundamental physics, is one of the mysteries of 
current science. 

The Pioneer anomaly is an intriguing clue in this context, and for this reason, 
already draws considerable attention in scientific press and beyond. 
A mission dedicated to the investigation of deep space gravity thus has a huge outreach potential 
which will be exploited by the mission team for disseminating scientific knowledge on the 
fundamental questions of modern gravitational physics, as well as on the instrumentation and 
space navigation techniques used to progress in this domain. 

As the ancient poem from which its name is drawn, Odyssey will consist in a 
long navigation to the borders of the explored world. 
During this journey, a number of events will take place in the Solar System, 
which the team has chosen to name after places or characters of the ancient poem: 
Sirens for the deep-space gravity test; Scylla and Charybdis for the investigation 
of fly-bys; Calypso for the long lasting Enigma stay in the outer Solar System; 
and Polyphemus the Cyclops for the solar conjunction tests.

\begin{acknowledgements}
Parts of the work were carried out at the Jet Propulsion Laboratory,
California Institute of Technology under contract with the National Aeronautics and
Space Administration and at Zarm and Astrium Satellites under the Enigma contract with the German Space Agency.
\end{acknowledgements}

\def\etal{\textit{et al }}
\def\ibid{\textit{ibidem }}
\def\url#1{\textrm{#1}}
\def\arxiv#1{\textrm{#1}}
\def\REVIEW#1#2#3#4{\textit{#1} \textbf{#2} {#4} ({#3})}
\def\Book#1#2#3{\textit{#1} ({#2}, {#3})}
\def\BOOK#1#2#3#4{\textit{#1} ({#2}, {#3}, {#4})}
\def\BOOKed#1#2#3#4#5{\textit{#1}, #2 ({#3}, {#4}, {#5})}
\def\Name#1#2{\textrm{#1}~#2}

\end{document}